# Modeling of magnitude distributions by the generalized truncated exponential distribution

Mathias Raschke

*Freelancer, Stolze-Schrey Str.1, 65195 Wiesbaden, Germany,*

*Mail: mathiasraschke@t-online.de, Tel.: +49 611 98819561*

Abstract

The probability distribution of the magnitude can be modeled by an exponential distribution according to the Gutenberg-Richter relation. Two alternatives are the truncated exponential distribution (TED) and the cut-off exponential distribution (CED). The TED is frequently used in seismic hazard analysis although it has a weak point: When two TEDs with equal parameters except the upper bound magnitude are mixed, then the resulting distribution is not a TED. Inversely, it is also not possible to split a TED of a seismic region into TEDs of sub-regions with equal parameters except the upper bound magnitude. This weakness is a principal problem as seismic regions are constructed scientific objects and not natural units. We overcome it by the generalization of the above-mentioned exponential distributions: the generalized truncated exponential distribution (GTED). Therein, identical exponential distributions are mixed by the probability distribution of the correct cut-off points. This distribution model is flexible in the vicinity of the upper bound magnitude and is equal to the exponential distribution for smaller magnitudes. Additionally, the exponential distributions TED and CED are special cases of the GTED. We discuss the possible ways of estimating its parameters and introduce the normalized spacing for this purpose. Furthermore, we present methods for geographic aggregation and differentiation of the GTED and demonstrate the potential and universality of our simple approach by applying it to empirical data. The considerable improvement by the GTED in contrast to the TED is indicated by a large difference between the corresponding values of the Akaike information criterion.

*distribution function, magnitude distribution, generalized truncated exponential distribution, AIC*

## 1. Introduction

The probability distribution of magnitudes is important for understanding and estimating seismic hazards and has been the object of several researches, including studies by Utsu (1999) and Kagan (2002). We intend to extend and improve the modeling of magnitude distributions. For this purpose, we briefly discuss important exponential distribution types and develop the generalized truncated exponential distribution (GTED) in the following section. We then discuss some aspects concerning the estimation of its parameters and apply a model to the data of the ISC catalog. Furthermore, we present possibilities for the geographic aggregation and differentiation. In the last section, we discuss the results and the questions arising. In this paper, we use the terminology of mathematical statistics; therefore, we generally refer to the dictionary of statistics of Upton and Cook (2008).

## 1. The exponential distributions

The well-known Gutenberg-Richter relation (Gutenberg and Richter, 1944) expresses the frequency with which a random magnitude *M* of an earthquake event exceeds a fixed level *m* by

$$N(m) = \alpha \exp(-\beta m).$$ (1)

It implies all earthquakes with random magnitude $M \geq m_{min}$ are exponentially distributed (Aki, 1965) with the cumulative distribution function $F_{expo}(m)$ and the corresponding survival function $\bar{F}_{expo}(m)$

$$F_{expo}(m) = 1 - exp(-\beta(m - m_{min}), M \geq m_{min} \text{ and}$$ (2a)

$$\bar{F}_{expo}(m) = 1 - F_{expo}(m) = exp(-\beta(m - m_{min})),$$ (2b)

wherein the survival function $\bar{F}_{expo}(m)$ describes the probability of exceedance *M>m*, and the cumulative distribution function $F_{expo}(m)$ describes the probability of non-exceedance *M≤m*. The exponential distribution has two advantages from the statistical point of view. First of all, it corresponds to the asymptotic tail distribution of many classic probability distributions, e.g. the normal or the gamma distribution (Beirlant et al. 2004, Tab. 2.3, Gumbel case) that provide a generalization. Secondly, the scale parameter *β* is independent of the definition of $m_{min}$ as indicated by equation (1). The formulation of equation (2b) also represents a normalized Gutenberg-Richter relation.

The Gutenberg-Richter relation and the corresponding exponential distribution have a disadvantage. They assume an infinite seismic energy. Furthermore, they often do not fit well to the observations in the range of larger magnitudes. The truncated exponential distribution with the survival function

$$\bar{F}_{tru}(m) = 1 - \frac{1 - exp(-\beta(m - m_{min}))}{1 - exp(-\beta(m_{max} - m_{min}))}, m_{max} \geq M \geq m_{min},$$ (3)

that partially overcomes these problems, has been introduced into seismology by Cosentino et al. (1977). There are also a number of statistical publications about the truncated exponential distribution, for example Hannon and Dahiya (1999). Utsu (1999) refers in his research on magnitude distribution models to more and earlier references from Cosentino et al. (1977). We could not examine everything, but Page (1968) only mentions the truncated exponential distribution in connection with the estimation of the scale parameter of the (un-truncated) Gutenberg-Richter relation. Furthermore, Utsu (1999) uses the correct probability density function (first [negative] derivation of our equation (2)) for the truncated exponential distribution. But he assigns it to the cut-off Gutenberg-Richter relation (his equation (9)), which corresponds to the cut-off exponential distribution. That is why we do not know exactly which reference of Utsu (1999) really deals with the truncated or the cut-off exponential distribution (CED). Kagan (2002) also considers the CED once in his research and refers to it as a characteristic distribution with survival function, being similar to equation (2b)

$$\bar{F}_{cut}(m) = exp(-\beta(m - m_{min})), \; m_{min} \leq M \leq m_{cut}. \tag{4a}$$

The jump in its graph (s. Fig. 1a or Kagan 2002, Fig. 1) at $m_{cut}$ generates the probability mass function

$$P(M = m_{cut}) = exp(-\beta(m_{cut} - m_{min})). \tag{4b}$$

The random magnitude with a cut-off distribution is a continuous and discrete random variable as its distribution function has continuous range for $M<m_{cut}$ and the discrete mass point at $M=m_{cut}$. The corresponding probability density function is only defined for the continuous range in contrast to the formulation of Kagan (2002, Eq. (6)).

The exponential distribution models have the aforementioned advantages in comparison to the exponential distribution. Nevertheless, in some cases they also do not fit very well to the larger magnitudes (see, e.g., Utsu 1999). Additionally, it has a principal weak point. When two TEDs or CEDs with equal parameters except the upper bound $m_{max}$ are mixed, then the resulting distribution is not a TED or CED. Inversely, it is also not possible to split a TED of a seismic region into two or more TEDs of sub-regions with equal parameters except the upper bound magnitude. This weakness is a fundamental problem as seismic regions are constructed scientific objects and not self-contained natural units. That is why a good and general distribution model for magnitudes should include the possibility of mixing and splitting.

We overcome this problem by developing the generalized truncated exponential distribution (GTED). Our basic idea is mixing exponential distributions with equal $\beta$ parameters by the simple product of the survival functions of this exponential distribution and the distribution of the random cut-off point $M_{cut}$ (in contrast to the fixed parameter $m_{cut}$). This is because all mixed CEDs with $M_{cut} > m$ are equal to the exponential distribution. We write

$$\bar{F}_{gen}(m) = \bar{F}_{exp}(m)\bar{H}(m), \ m_{min} \leq M < m_{max}, \tag{5a}$$

that implies for the range $m \leq m_d$ with $m_d \leq m_{max}$

$$\bar{F}_{gen}(m) = \bar{F}_{exp}(m), \ m_{min} \leq M < m_d. \tag{5b}$$

The survival function of the cut-off points has the characteristic outside the interval $(m_d, m_{max})$ that

$$\bar{H}(m) = 1, m \leq m_d \quad \text{and} \tag{6a}$$

$$\bar{H}(m) = 0, m \geq m_{max}. \tag{6b}$$

Note that the approach works not only for the exponential distribution but also for several other types of continuous distribution functions. The probability density function is the negative first derivation of the survival function

$$f_{gen}(m) = -\frac{d\bar{F}_{gen}}{dm} = f_{exp}(m)\bar{H}(m) + \bar{F}_{exp}(m)h(m), \ m_{min} \leq M < m_{max}, \tag{7}$$

where $f_{exp}(m)$ stands for the probability density function of the exponential distribution and $h(m)$ is the probability density function of $M_{cut}$. This distribution can be modeled by any distribution type and influences the properties of the GTED, such as its characteristic function and its moments. In all cases, $m_d$ must not be smaller than $m_{min}$. The upper bound $m_{max}$ of the survival function $\bar{H}(m)$ equals that of the overall GTED. The density function $f_{gen}(m)$ has a jump at $m = m_d$ if $h(m_d) > 0$. Additionally, the GTED has the basic properties that $\bar{F}_{gen}(m) \leq \bar{F}_{cut}(m)$ and $\bar{F}_{gen}(m) \leq \bar{H}(m)$. Furthermore, the GTED is a generalization with the special cases CED, TED and exponential distribution. The CED is formulated by the GTE with $m_d = m_{max}$ being finite. When $m_d = m_{max}$ is infinite, the GTED is equivalent to the classical exponential distribution; and the truncated exponential distribution is a GTED with $\bar{H}(m) = \bar{F}_{tru}(m)/\bar{F}_{exp}(m)$ and $m_d = m_{min}$.

Any distribution can be used for $\bar{H}(m)$, the bounds only have to be $m_{max} \geq m_d \geq m_{min}$. But the upper bound $m_{max}$ could be infinite. We suggest the beta distribution for the modeling of $\bar{H}(m)$, as it is very flexible. Its cumulative distribution function and its survival function cannot be expressed explicitly for this distribution type, but can be computed numerically (e.g. by a worksheet function of MS Excel). The probability density function of the beta distribution for the random variable $X$ and the real number scale $x$ is the following:

$$f_{beta}(x) = \frac{\left(\frac{(a-x)}{a-b}\right)^{c-1}\left(1-\frac{(a-x)}{a-b}\right)^{d-1}\Gamma(c+d)}{(a-b)\Gamma(c)\Gamma(d)}, a \leq X \leq b, c > 0, d > 0, \qquad (8)$$

where the parameters $a$ and $b$ define the interval of $X$, while $c$ and $d$ are shape parameters. The elementary gamma function is outside of the range between $a$ and $b$, $f_{beta}(x)$=0. $\Gamma(.)$ . For further details, see Johnson et al. (1995, section 25).

We present some examples of the GTED in Fig. 1a with mixing beta distributions. Therein, we also show a TED, CED and exponential distribution, that are only special cases of the GTED. The flexibility of the GTED is obvious. The corresponding functions $\bar{H}(m)$ are represented in Fig. 1b.

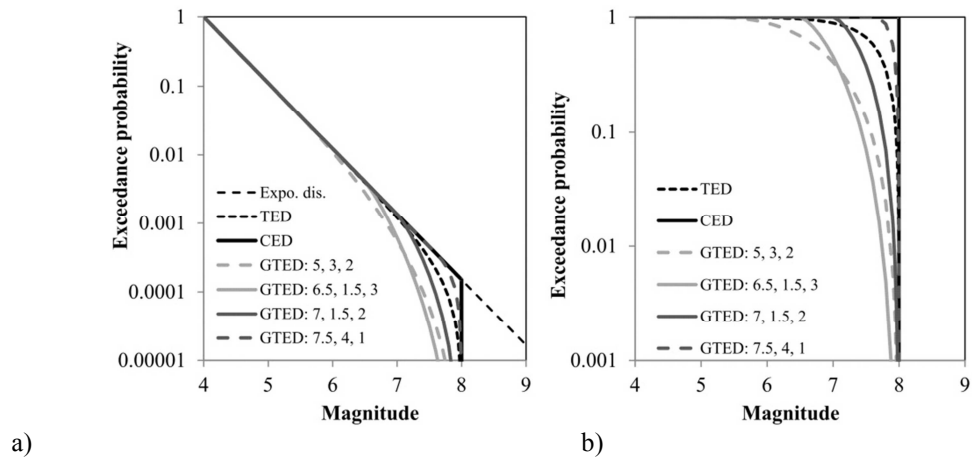

Figure 1: Examples of the GTED with beta distributed cut-off points for $m_{max}$=8: a) survival functions also for the special cases TED, CED and exponential distribution, b) corresponding survival functions $\bar{H}(m)$ of the cut-off points except for the exponential distribution with infinite $m_c$ (parameter values in the legend: $m_d$, $c$ and $d$).

# 2. An empirical example and aspects of the parameter estimation

We discuss here the issue of parameter estimation for the GTED using the example of the moment magnitudes of earthquakes included in the catalog of the International Seismological Centre (ISC). The earthquakes occurred between 1970 and 2009, considering an $m_{min}$=5,595. We use the entire catalog without differentiation in order to demonstrate the universality of the approach. Furthermore, we have the largest possible number of large events with $M > m_d$ for the entire catalog and we apply the Akaike information criterion (Akaike, 1974; Lindsey, 1996; Upton and Cook, 2008; AIC) to compare different alternative models objectively. This criterion works better for larger sample sizes (e.g. Acquahns, 2010; Raschke and Thürmer, 2010); and so does the estimation of the upper bound $m_{max}$.

The magnitudes of our sample are listed with a precision of two decimal places. The empirical distribution function of this sample is computed for the ordered sample $M_1 \leq M_2 \leq ... \leq M_i \leq ... \leq M_n$ by $\bar{F}_i = 1 - 1/(n+1)$ with sample size $n$=12,475. It is represented in Fig. 2.

We can estimate the $\beta$ parameter of the exponential part of the GTED by the normalized spacing. When we have an exponential distributed random variable $M$ with the ordered sample $M_1 \leq M_2 \leq ... \leq M_i \leq ... \leq M_n$, then we also have an exponential random variable $Y_i = (n+1-i)(M_i - M_{i-1})$ with the independent observations and the same distribution parameter $\beta$ according to Sukhatme (1936). We consider here all $Y_i$ for which $M_i < m_d$ in order to estimate $\beta$ from the reciprocal of the sample average of $Y$, in accordance with Deemer and Votaw (1955). The actual estimation can be done iteratively with the current estimation of $m_d$. In our example, $Y$ is not really exponentially distributed due to the limited number of decimal places in the data, but this does not seriously affect the estimation of $\beta$.

We can also use the normalized spacing to check the completeness of the earthquake catalogue. We split the sample of $Y_i$ into two even sub-samples: one for the smaller magnitude with $m_{min} < M_i \leq m_{split}$ and the other with $m_{split} < M_i \leq m_d$. If there is a significant lack of events of smaller magnitude in the catalog, then the sample mean of $Y_i$ of the smaller magnitudes should be significantly larger than for the larger magnitudes. The reason for this is that the absolute value of slope of the logarithmized survival function should be smaller in the range of smaller, incomplete magnitudes. The significance of the differences between the two sample means can be tested by the classical t-test (student test) for two samples. The difference between the two sample means of our data with $\hat{m}_d$=7.395 and $m_{split}$=5.895 is not significant for a significance level larger than 10%; any incompleteness is not relevant.

There are different estimation methods for the upper bounds of a truncated distribution including the corresponding confidence regions (Kijko and Singh, 2011; Raschke, 2012). We apply the simple method by Robson and Whitlock (1964) with the estimation $\hat{m}_{max} = M_n + (M_n - M_{n-1})$, where the two largest observations are $M_n$ and $M_{n-1}$. This estimator does not need any further information or pre-estimation. Successively, we estimate all other parameters of the mixing beta distribution $m_d$, $c$ and $d$ with the well-known maximum likelihood estimation. Therein, the well-known logarithmized likelihood function

$$ln(L) = \sum_{i=1}^{n} ln\left(f_{gen}(M_i; m_d, c, d)\right)$$ is maximized (see, e.g., Coles, 2001,

section 2.6.3). The estimation error (standard error) can be computed by applying the Fischer information matrix (see, e.g., Coles, 2001, section 2.6.4), which has already been applied in seismology (e.g. Rhoades, 1997, his equation (15) for a regression model). For our example, we have estimated the parameters $\hat{m}_d$=7.395±0.012, $\hat{c}$=1.594±0.243 and $\hat{d}$=3.132±0.573. There are 195 observations with $M > m_d$ in the tail, which is much more than for many other distribution models.

The scale parameter of the exponential part is $\hat{\beta}$=2.308±0.020. Its estimated standard error, obtained from the Fischer information matrix, is equal to the classical estimation with $\hat{\beta}/\sqrt{n}$ according to Deemer and Votaw (1955). The estimated upper bound is $\hat{m}_{max}$=9.380±0.380 according to Hannon and Dahiya (1999).

We show the estimated GTED in Fig. 2 and compare it with the exponential, truncated exponential and cut-off exponential distributions.

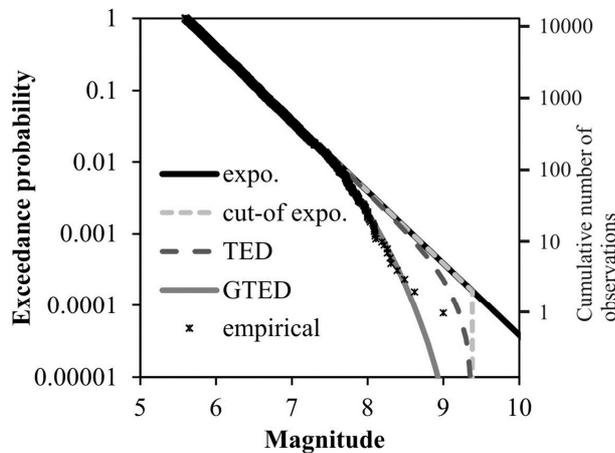

Figure 2: Estimated exponential distributions for the sample of the ISCcatalog

We can easily observe in Fig. 2 that the GTE fits much better, nevertheless, we can objectively compare the distribution models using the aforementioned $AIC=-2ln(L)+2p$ (Akaike, 1974; Lindsey, 1994; Upton and Cook, 2008) with the aforementioned logarithmized likelihood function $ln(L)$. $p$ is the number of estimated parameters. The smallest AIC detects the best model. The main idea of this criterion is that a larger likelihood value indicates a better model but a larger number of estimated model parameters $p$ should be penalized. Otherwise, we would cause over-parameterization (over-fit). The AIC works similarly to the likelihood ratio test, its functionality has been validated by numerical researches (e.g. Acquahns, 2010; Raschke and Thürmer, 2010) . The AIC is very popular in mathematical statistics, applied to various statistical models such as magnitude distributions (Utsu 1999), regression models (e.g. Rawlings et al. 1998) or copulas (e.g. Genest et al., 2007).

The ordinary exponential distribution has an AIC=4118.48 with $p$=1 in our example. The TED has an AIC=4116.31 with $p$=2. The GTED shows the best fit, with an AIC=4102.28 and $p$=5. The improvement of the magnitude distribution model by the TED is powerful; even if we applied $p$=10 parameters, the model would be the best with AIC=4112.28.

We would discourage applying the Bayesian information criterion (BIC) from Schwarz (1978) here for the model selection, as the value of $m_{min}$ would considerably influence the sample size and, through this, the model selection with the BIC. Therein, the sample size for the tail with $M>m_d$ would be fixed, although the entire sample size $M \geq m_{min}$ is changed. This situation is not considered appropriately by the BIC.

We have estimated the magnitude distribution for the entire data set under the assumption that all magnitude realizations are independent of each other. This assumption is not altered automatically by any clustering in the seismicity, for example aftershock activity, as only the occurrence intensity is influenced by the main shock, not the random realization of the aftershock magnitude in itself. When we consider the exponential model with equation (1) and (2), a main shock can modify parameter $\alpha$ in equation (1) and consequently change the number of earthquake events with $M \geq m_{min}$ within the following time period. On the other hand, the random magnitudes of this time period are independent realizations of the exponential distribution according to equation (2). Otherwise, space-time models such as those by Reasenberg and Jones (1989) or Ogata (1998) would explicitly include a conditional probability distribution for the aftershock magnitudes. This is not the case. There is a misunderstanding concerning the randomness of the earthquake process in publications that demands a sort of de-clustering in order to insure "uncorrelated magnitudes" (e.g. Holschneider, 2011).

A further consideration has to be made about $m_d$. The estimated value implies that there is no region on Earth with an upper bound magnitude smaller than 7.395. Of course, this is unlikely. This can be explained by the imperfection of our model and the very limited occurrence of such regions in the entire seismicity on Earth. Nevertheless, our distribution model fits the observations very well according to Fig. 2. There is no statistical indication that the observations of the entire ISC catalog could not be modeled by one GTED. This does not exclude spatial and/or temporal differentiation for the concrete application of a GTED. The AIC could also be used for finding a limit of reasonable differentiation of the magnitude distribution according to the geographic region or source process. Otherwise, there is the danger of over-parameterization (over-fit).

## 3. Geographic aggregation and differentiation

The GTED of $k$ regions can easily be aggregated if the parameter $\beta$ is the same for all regions. The distribution of the cut-off point has only to be aggregated as

$$\bar{H}_{all}(m) = \frac{\sum_{i=1}^{k} \Lambda_i \bar{H}_i(m)}{\sum_{i=1}^{k} \Lambda_i}, \qquad (9)$$

where $\Lambda_i$ is the occurrence frequency of region $i$ for all events with $M \geq m_{min}$; $m_{max\_all}$ is the maxima of all $m_{max\_i}$, and $m_{d\_all}$ is the minima of all $m_{d\_i}$. A split of one TED into two or more TEDs can be done by an inverse procedure.

Furthermore, we can understand the magnitude distribution of the GTED as the aggregation of the magnitude distributions of all potential source points **s** (coordinate vectors) in a region, or at a fault line, with fixed parameter $\beta$. Therein, the magnitude distribution of each source point **s** is a cut-off exponential distribution with a fixed cut-off point $m_{cut}(\boldsymbol{s})$. The random cut-off point is generated by a random realization of the source point **s** with $M_{cut} = m_{cut}(\boldsymbol{s})$. The corresponding survival function $\bar{H}(m)$ of the entire region can be computed by the occurrence density $\lambda(\boldsymbol{s})$ for $M \geq m_{min}$ and with the indicator function $\mathbf{1}(\boldsymbol{s}, m)$

$$\bar{H}(m) = \frac{\int \mathbf{1}(\mathbf{s}, m)\lambda(\mathbf{s})d\mathbf{s}}{\int \lambda(\mathbf{s})d\mathbf{s}}, with \ \mathbf{1}(\mathbf{s}, m) = 1 \ if \ m \leq m_{cut}(\mathbf{s}),$$

$$otherwise \ \mathbf{1}(\mathbf{s}, m) = 0. \tag{10}$$

This approach can also be applied inversely to differentiate the magnitude distribution of an entire region for each source point **s**, depending on a geographic quantity y(**s**), when $\bar{H}(m)$ is known. The survival function $\bar{F}_y(m)$ has to be computed as

$$\bar{F}_y(m) = \frac{\int \mathbf{1}(\mathbf{s},m)\lambda(\mathbf{s})d\mathbf{s}}{\int \lambda(\mathbf{s})d\mathbf{s}}, with \ \mathbf{1}(\mathbf{s}, m) = 1 \ if \ m \leq y(s),$$

$$otherwise \ \mathbf{1}(\mathbf{s}, m) = 0. \tag{11}$$

The site-related cut-off points $m_{cut}(\mathbf{s})$ can then be computed by the equation

$$\bar{H}(m_{cut}(\mathbf{s})) = \bar{F}_y(y(\mathbf{s})). \tag{12}$$

The geographic quantity y(**s**) could also be the occurrence density $\lambda(\mathbf{s})$. The occurrence frequency $\Lambda$ for the entire region or fault line for all $M \geq m_{min}$ is $\int \lambda(s)ds$. A time-related differentiation is also conceivable.

## 4. Conclusion, Discussion and Outlook

We have developed the GTED as a simple mixing of the exponential distribution and the distribution of corresponding cut-off points. It is a generalization that includes the ordinary exponential distribution, the CED and the TED as special cases. The GTED overcomes the weak point of TEDs that is one TED cannot be divided into two TEDs with different upper bounds and cannot be the result of the mixing of two TEDs. A GTED can be aggregated from a number of GTEDs and de-aggregated to GTEDs if the parameter $\beta$ is the same for all. This simplicity is a further advantage of the new approach, besides the fact that the GTED is equal to the exponential distribution in the range of smaller magnitudes. Its flexibility is also an advantage; we can use any distribution model for the cut-off points. The beta distribution is only a suggestion and has been used here to demonstrate the opportunities of the GTED.

Beside this, we have presented suitable estimation methods for the parameters of the GTED. The normalized spacing is introduced here for estimating the parameter $\beta$ and for examining the completeness of the sample. Furthermore, we have demonstrated the advantages of the proposed methodology by estimating the magnitude distribution of a sample from the ISC catalog. The AIC of the GTED is much smaller than the AICs of the TED and the exponential distribution, which indicates a very strong improvement of the modeling of the magnitude distribution.

The GTED is a good tool for modeling the distribution of earthquake magnitudes. Nevertheless, we see a number of questions remain open. How could the distribution model be tested? The $\text{Chi}^2$ test is a possibility, but it is not powerful enough (see, e.g., Raschke, 2009). The behavior of the estimators of the upper bound magnitude should also be investigated in future research. We expect that the classical estimation methods of the extreme value statistics (e.g. Coles, 2001) will not work as well as for the truncated exponential distribution (Raschke, 2012). Similarly, the asymptotic behavior and the corresponding extreme-value indices of the GTED should also be further investigated (see, e.g., Beirlant et al., 2004). Besides this, the estimation of the special case $m_{min}=m_c$ should be analyzed; we expect some problems as different parameter combinations could result in very similar GETDs. The geographic differentiation could also be considered within more complex estimation procedures.

Beside these, we have only presented a model. If we would consider further observations, then the parameter estimation would change. But this applies to all models. And if we would consider a smaller geographic region or a certain type of earthquake, then parameter $\beta$ would be different. This also applies to the TED or the exponential distribution. In such cases, the criteria of statistical model selection (e.g. AIC and BIC) could help in finding an appropriate and objective limit for the differentiation. But this issue is not the topic of our research.

Finally, it should not be forgotten that what cannot be improved in any way by statistical models and methods is the accuracy of the magnitude measurements.

## Data and Resources

The earthquake catalog of the International Seismological Centre (ISC, *www.isc.ac.uk/iscgem/download.php*, version 1.04, released on 2013-11-05, period covered: 1900-2009; last accessed in February 2014).